\documentclass{article}
\usepackage[preprint]{spconf}
\usepackage{amsmath,graphicx, amsfonts,,amsthm,amssymb,enumerate,enumitem,mathtools,float,accents,hyperref,url,enumerate, xcolor, makecell, booktabs, caption, subcaption}

\copyrightnotice{\raisebox{-5pt}{\parbox{\textwidth}{\copyright\ 2024 IEEE. Personal use of this material is permitted. Permission from IEEE must be obtained for all other uses, in any current or future media, including reprinting/republishing this material for advertising or promotional purposes, creating new collective works, for resale or redistribution to servers or lists, or reuse of any copyrighted component of this work in other works.}}}

\graphicspath{ {./figures/} }

\setlist[itemize]{noitemsep}

\title{Towards an interpretable representation of speaker identity via perceptual voice qualities}
\makeatletter
\def\@name{
  \emph{Robin Netzorg}$^{1}$,
  \emph{Bohan Yu}$^{1}$,
  \emph{Andrea Guzman}$^{1}$, \\
  \emph{Peter Wu}$^{1}$,
  \emph{Luna McNulty}$^{2}$,
  \emph{Gopala Anumanchipalli}$^{1}$
}
\makeatother
\address{
  $^1$University of California, Berkeley,
  $^2$Brown University
}
\begin{document}
%\ninept
%
\maketitle
\begin{abstract}
Unlike other data modalities such as text and vision, speech does not lend itself to easy interpretation. While lay people can understand how to describe an image or sentence via perception, non-expert descriptions of speech often end at high-level demographic information, such as gender or age. In this paper, we propose a possible interpretable representation of speaker identity based on perceptual voice qualities (PQs). By adding gendered PQs to the pathology-focused Consensus Auditory-Perceptual Evaluation of Voice (CAPE-V) protocol, our PQ-based approach provides a perceptual latent space of the character of adult voices that is an intermediary of abstraction between high-level demographics and low-level acoustic, physical, or learned representations. Contrary to prior belief, we demonstrate that these PQs are hearable by ensembles of non-experts, and further demonstrate that the information encoded in a PQ-based representation is predictable by various speech representations. 
\end{abstract}
%

% We leave the application of the PQ-based representations to future work. 

\begin{keywords}
Speech Representation, Speaker Identity, Perceptual Qualities
\end{keywords}
\section{Introduction}
\label{sec:intro}

When a lay person hears a voice, they can quickly identify certain features, such as whether or not the voice is masculine or feminine, or old or young \cite{schweinberger2014impression}. This level of abstraction, however, does not shed light on the building blocks of a voice. Intermediate representations of speech are seemingly locked behind a veil of expertise, time-consuming to understand and contained to sub-fields. While there have been certain combinations of various fields and speech processing, such as speech pathology and musical vocal training, these combinations have been disconnected and not applied towards creating a complete representation of speaker identity. 

In this work, we holistically consider one such intermediate representation from speech pathology: perceptual voice quality. Prior work has explored the ability of crowdsourcing and machine learning methods to capture these individual qualities \cite{hidaka2022deepgrbas, mcallisterperceptual2023}, but, to the best of our knowledge, a study of perceptual qualities' ability to represent speaker identity has not been conducted. Perceptual qualities related to voice atypicality alone lack the ability to provide a comprehensive latent space of adult voices, since gender information is missing. As such, we take inspiration from musical and transgender vocal training, supplementing perceptual qualities from speech pathology with what we call gendered perceptual qualities. Combined with perceptual qualities from the Consensus Auditory-Perceptual Evaluation of Voice (CAPE-V) protocol \cite{capev2009}, we propose a 7-dimensional representation of spoken speaker identity based on perceptual qualities. 

% Firstly, while not all perceptual qualities are included in this work, the perceptual qualities listed here provide a complete space of speaker identity, wherein any speaker can be mapped to a point in this space, with a high degree of uniqueness (\robbie{idk if we can claim this}). 

A perceptual quality-based representation of speaker identity provides two benefits currently lacking in other representations. Firstly, a perceptual quality-based representation of speaker identity is low-dimensional and interpretable, whereby any listener or ensemble of listeners, given minimal training, can hear the particular aspects of each perceptual quality. Secondly, the information encoded in subjective perceptual qualities has an objective basis, containing information present in varied representations of speech, from hand-crafted to self-supervised. Perceptual qualities have the potential to bring to speech processing what it has previously lacked: a perceivable and descriptive level of abstraction of the texture of a speaker's voice. 

% Our contributions are as follows:
% \begin{itemize}
%     \item Addition of voice teacher labellings of gendered perceptual qualities to the PVQD.
%     \item Evidence of the ability of non-experts to accurately label perceptual qualities, which has historically relied on the knowledge of trained experts.
%     \item An exploration of the connection between multiple perceptual qualities and modern and varied representations of speech. 
% \end{itemize}

\section{Relation to Prior Work}
\label{sec:related}

\subsection{Perceptual Voice Quality}

Defined as the acoustic "coloring" of an individual's voice, perceptual voice quality, hereby referred to as perceptual quality (PQ), has long been studied in speech language pathology and processing \cite{kreiman1993perceptual, kellervoicequality2005}. From the mood of a voice to vocal fry and breathiness, perceptual quality consists of the subjective perceptions of a voice. Perceptual qualities have long been noted as being important to spoken language processing, with prior work noting that voices with uncommon or pathological perceptual qualities lead to poor performance for spoken language processing systems if not taken into consideration \cite{kellervoicequality2005}. 

Of particular interest to our work is how perceptual quality can be used to describe an individual's voice. In speech language pathology (SLP), experts will use vocal quality to perform initial diagnosis of the health of an individual's voice \cite{capev2009, pqvd2020}. Non-surgical treatment of a voice involves a patient performing exercises to bring certain perceptual qualities into healthy levels \cite{capev2009}. PQs that are highly correlated with dysphonic speech are particularly useful for voice rehabilitation, but others, such as vocal fry, timbre/resonance, and weight, see application in general voice modification as well. Musical vocal teachers, 
 SLPs or voice teachers specializing in voice feminization/masculinization will use these other PQs to guide students towards target voices \cite{diamant2021examining, carew2007effectiveness}. 

Prior work in automatic assessment of dysphonia \cite{garcía2022deep} and emotion recognition \cite{zhou2022emotion} have explored the predictability of individual perceptual qualities. Deep neural network approaches to predicting perceptual qualities have been conducted \cite{hidaka2022deepgrbas, fujimura2022classification}, but these have primarily focused on labeled audio clips of sustained vowels with the goal of predicting dysphonic voices from spectral features and the waveform directly. We extend on prior work by demonstrating that automatic detection of perceptual qualities is possible at a human level across multiple representations of spoken sentences.

% Describing a voice can be a difficult task, but a complete mapping of perceptual qualities alleviates this. Prior work in controllable voice synthesis has primarily focused on prosody control \cite{hu2022prosodybert}, or other primarily focusing on a single perceptual quality, like emotion \cite{zhou2022emotion}. Voice conversion methods will shift perceptual qualities, but often do so through end-to-end training and do not allow for fine-grained modification of a voice \cite{lian2022robust}. In this work, we aim to produce a voice modification system that takes in multiple forms of perceptual qualities, and allows for control over these perceptual qualities.

\subsection{Representations of Speaker Identity}

From speaker verification and identification \cite{ivector, xvector} to voice conversion \cite{lian2022robust}, an informative representation of speaker identity is necessary for many problems in speech processing. Across nearly all modern methods, especially those based in deep learning \cite{bai2021speaker}, the highest performing methods are learned representations that, while containing information relevant to speaker identity, lack highly low-dimensional interpretability of the PQ-based representation proposed here.

\section{Perceptual Voice Qualities}

% \begin{table}[]
%     \centering
%     \begin{tabular}{|c|c|}
%         \hline
%          \textbf{Perceptual Quality} & \textbf{Description}  \\
%         \hline
%          Resonance &  \makecell{Sound quality of the \\size of the vocal tract} \\
%          \hline
%          Weight & \makecell{Sound quality of the\\ vocal fold vibratory mass} \\
%          \hline
%          Strain &  \makecell{Perception of excessive\\ vocal effort (hyperfunction)} \\
%          \hline
%          Loudness & Deviation in loudness \\
%          \hline
%          Roughness & \makecell{Perceived irregularity \\in the voicing source} \\
%          \hline
%          Breathiness &  Audible air escape in the voice \\
%          \hline
%          Pitch & Deviation in pitch\\

%         \hline
%     \end{tabular}
%     \caption{The list of perceptual qualities studied and their descriptions. Resonance and weight are taken from vocal training, and the rest are taken from the CAPE-V protocol.}
%     \label{tab:pvq}
% \end{table}

In this section, we describe the collection and interpretability of perceptual voice qualities. While the space of possible PQs is vast, we limit our consideration to seven PQs, described below. 

\subsection{Perceptual Voice Qualities Database}

In clinical settings, perception assists greatly in the early stages of diagnosis of voice pathologies. Speech Language Pathologists (SLPs) will often use rating scales like the Consensus Auditory-Perceptual Evaluation of Voice (CAPE-V) to provide early information on possible voice pathologies an individual may have \cite{capev2009}. SLPs undergo training to successfully identify the PQs: strain, loudness, roughness, breathiness, pitch, and severity \cite{pqvd2020}. For a complete description of the CAPE-V vocal qualities, please refer to the original protocol ~\cite{capev2009}.

While this data is difficult to collect, the Perceptual Voice Qualities Database (PVQD) serves a publicly available ratings of PQs from the CAPE-V scale \cite{pqvd2020}. The PVQD includes 296 audio files of around 30 seconds of audio, whereby a speaker follows the CAPE-V evaluation protocol, and reads six sentences and produces vowels $/a/$ and $/i/$ for 1-2 seconds. The authors behind the PVQD had each audio clip rated by three separate clinicians across two trials according to the CAPE-V scale. In this work, we examine five of the six PQs, excluding severity, which is overall measure of vocal atypicality.

\begin{figure*}[t]
     \centering
     \begin{subfigure}[b]{0.22\textwidth}
         \centering
         \includegraphics[width=\textwidth]{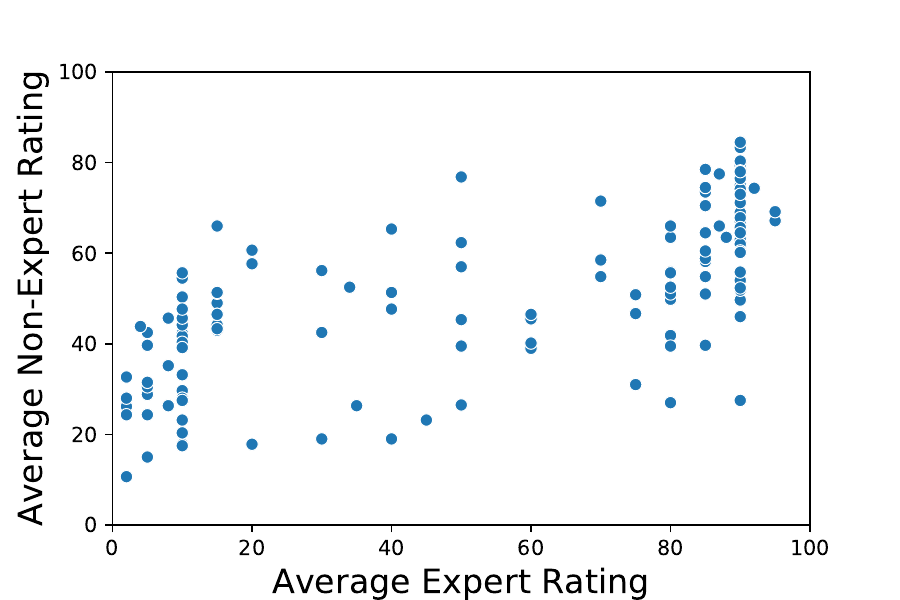}
         \caption{Resonance}
         \label{fig:ne_res}
     \end{subfigure}
     \begin{subfigure}[b]{0.22\textwidth}
         \centering
         \includegraphics[width=\textwidth]{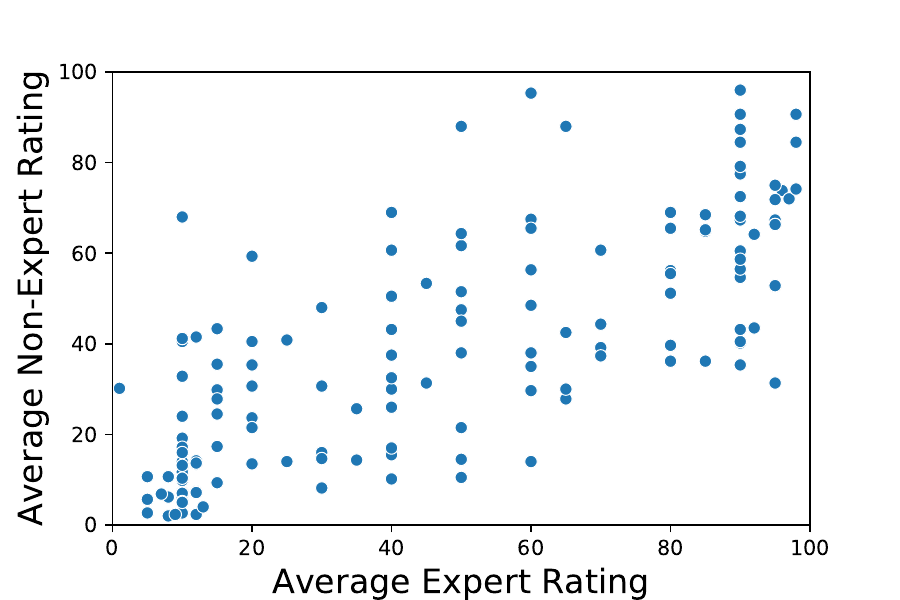}
         \caption{Weight}
         \label{fig:ne_weight}
     \end{subfigure}
     \begin{subfigure}[b]{0.22\textwidth}
         \centering
         \includegraphics[width=\textwidth]{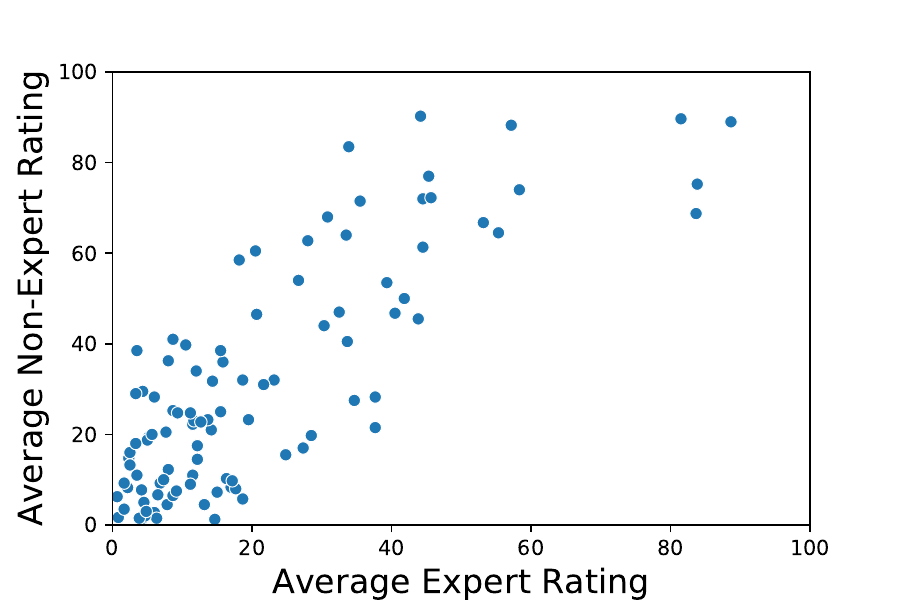}
         \caption{Strain}
         \label{fig:ne_strain}
     \end{subfigure}

     \begin{subfigure}[b]{0.22\textwidth}
         \centering
         \includegraphics[width=\textwidth]{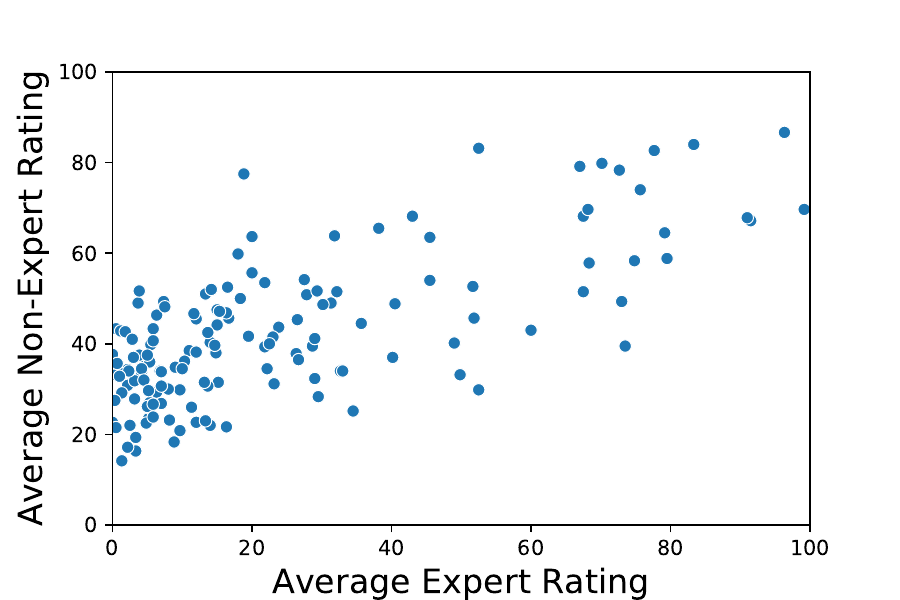}
         \caption{Loudness}
         \label{fig:ne_loudness}
     \end{subfigure}
     \hfill
     \begin{subfigure}[b]{0.22\textwidth}
         \centering
         \includegraphics[width=\textwidth]{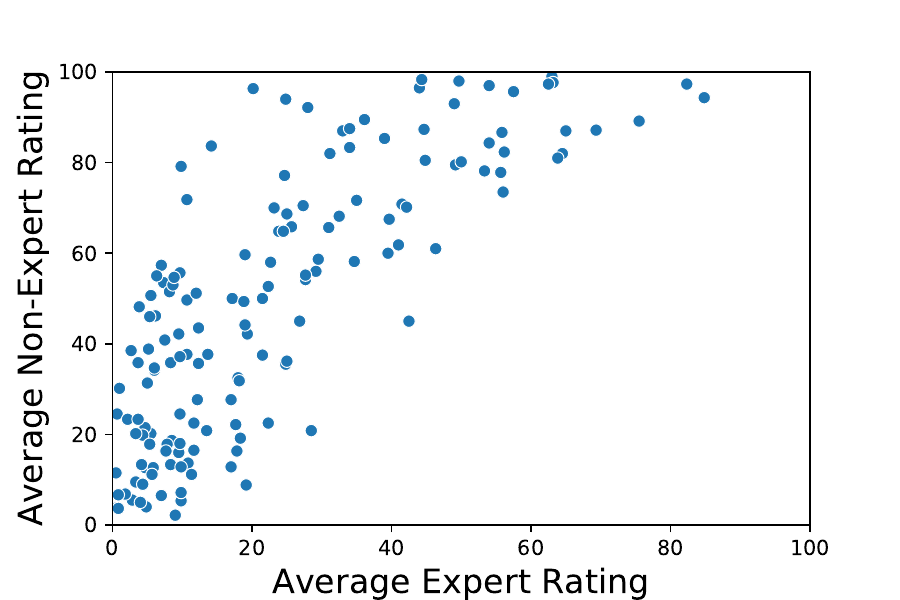}
         \caption{Roughness}
         \label{fig:ne_roughness}
     \end{subfigure}
     \hfill
     \begin{subfigure}[b]{0.22\textwidth}
         \centering
         \includegraphics[width=\textwidth]{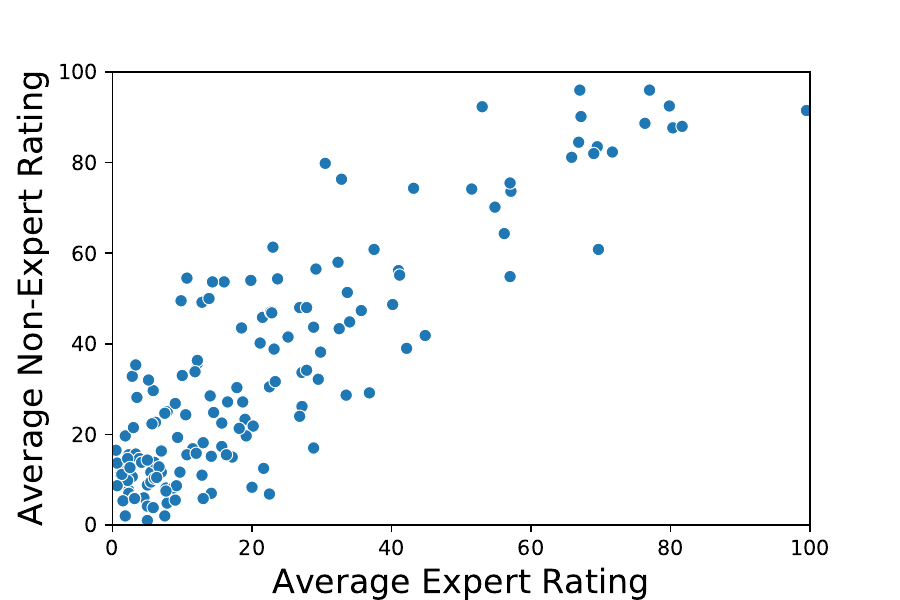}
         \caption{Breathiness}
         \label{fig:ne_breathiness}
     \end{subfigure}
     \hfill
     \begin{subfigure}[b]{0.22\textwidth}
         \centering
         \includegraphics[width=\textwidth]{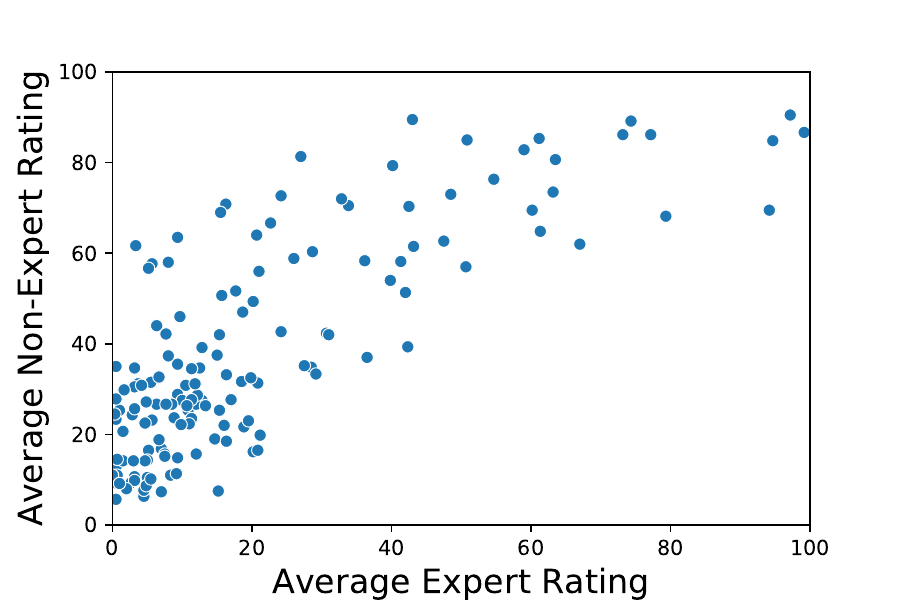}
         \caption{Pitch}
         \label{fig:ne_pitch}
     \end{subfigure}
        \caption{Average Expert Rating (x-axis) vs. Average Non-Expert Rating (y-axis) across perceptual qualities.}
        \label{fig:ne_vs_e}
\end{figure*}

\subsection{PVQD+: Collecting Gendered Perceptual Qualities}\label{sec:pvqd_plus}

While the PVQD serves its purposes as a diagnostic tool of speech pathology, for the purposes of providing a general representation of voice, it is incomplete. Information on a voice's gender measures of is missing. Attempting to perform manipulations that are common in speech processing tasks like voice conversion, such as converting a masculine voice to a feminine voice, would not be possible with the  CAPE-V scale's ratings of deviation alone. 

As such, we augment the labels with those provided by three voice teachers, who specialize in transgender voice training. 
Current pedagogies in transgender voice modification
are particularly concerned with two primary PQs:
vocal resonance and vocal weight.
These correspond to two physiological differences \cite{markova2016age}
that distinguish male and female voices.
Perceptual resonance corresponds to the amount of space 
above the vocal folds in the vocal tract.
More space causes lower resonant frequencies to be amplified \cite{kent1993vocal},
resulting in a deeper timbre, even at high pitches.
Perceptual weight corresponds to the vibratory mass of the vocal folds,
which is correlated with the open quotient,
(the proportion of the glottal cycle during which the vocal folds are open)
and spectral slope 
(the decline in amplitude from the first to the Nth harmonic) \cite{zhang2016cause}.
These ideas are also often used in singing lessons for vocalists, but the focus of transgender voice training on gender lines lends itself to a representation of spoken voice.

% Such lessons often involve individuals changing the resonance (timbre resulting from the amount of space in the vocal tract) and weight (vocal mass) to match those of the patient's assigned gender. 

As discussed above, three voice teachers listened to the subsets of 100 audio clips from PVQD and provided a label of resonance and weight on a scale 1-100. For resonance, a value of 1 represented the darkest resonance possible and a value of 100 represented the brightest resonance possible. Similarly for weight, a value of 1 represented the lightest voice possible and a value of 100 represented the heaviest voice possible. Healthy feminine voices were given a resonance value of 90 and a weight value of 10, and the opposite for healthy masculine voices. We note that one voice teacher labeled the entire dataset, and the two other voice teachers overlap on 25 audio clips to allow for the calculation of averages and correlations. We call the extended dataset PVQD+. 

% We provide a summary of the final perceptual qualities in Table \ref{tab:pvq}.

Along with those reported in the original PVQD, we report the intra-class correlation (ICC) for expert ratings of resonance and weight in Table \ref{tab:cor}. ICC is a common metric for measuring inter-rater reliability \cite{pqvd2020}. We see that the expert ICC of both resonance and weight are the maximum and minimum of the ICCs for all PQs, respectively. We note that the high inter-rater reliability for resonance is unsurprising, given that vocal tract size is highly correlated with gender. Weight is less clear. While an ICC of 0.77 still suggests high inter-rater agreement, weight having the lowest agreement among experts suggests that labelling weight is a more difficult task. Regardless, these results demonstrate the high agreement of experts on the gendered perceptual qualities resonance and weight.

\begin{table*}[t]
    \centering
    % \begin{tabular}{*{8}{c}}
    \begin{tabular}{|c|c|c|c|c|c|c|c||c|}
        \hline
        \textbf{Rater} & \textbf{Resonance} & \textbf{Weight} & \textbf{Strain} & \textbf{Loudness} & \textbf{Roughness} & \textbf{Breathiness} & \textbf{Pitch} & \textbf{Average}  \\
        \hline
         Non-Experts & 0.68 & 0.76 & 0.81 & 0.67 & 0.79 & 0.87 & 0.79 & 0.77 \\
         \hline
         Experts & 0.91 & 0.77 & 0.83 & 0.87 & 0.79 & 0.83 & 0.86 & 0.84 \\

         \hline
    \end{tabular}
    \caption{Correlation of Non-Expert Labels with Expert Labels. A measure of inter-rater agreement, Intra-class Correlation Coefficient, is reported for Expert Labels.}
    \label{tab:cor}
\end{table*}

\subsection{Can Non-Experts Hear Perceptual Qualities? }

A common statement from voice teachers and speech language pathologists is that hearing perceptual qualities requires training \cite{nagle2022capevclinic, huff2022modern}. The obstacle of expertise calls into question the utility of perceptual qualities as a representation of speaker identity, since collecting additional labels would be a costly and time-consuming task \cite{hidaka2022deepgrbas}. Recent work has demonstrated that non-experts can accurately label the overall atypicality of a voice \cite{mcallisterperceptual2023}, but no work has explored the ability of non-experts to label specific perceptual qualities. If non-experts can accurately label PQs, collecting a large-scale and high-quality dataset of perceptual qualities and using perceptual qualities as an evaluation metric of speaker modification systems would be possible. 

In this section, we test the ability of non-experts to accurately rate the PQs of PVQD voice clips. Using the Amazon Mechanical Turk (AMT), we ask 6 workers with master's qualifications to rate the clips using the CAPE-V protocol, and the resonance and weight as described in Section \ref{sec:pvqd_plus}. Workers are provided two examples for each perceptual quality, one example being low in that quality (ex. No Strain) and the other being high in that quality (ex. High Strain). Due to cost constraints, workers labeled only 150 audio clips of the 296 audio clips in PVQD. 

The results of the AMT experiments are very promising for the ability of non-experts to hear perceptual qualities. As Table \ref{tab:cor} demonstrates, average non-expert ratings achieve a correlation 0.77 with average expert ratings, with the lowest correlations being 0.68 and 0.67, for resonance and loudness respectively (loudness was often conflated with audio clip volume, not inherent loundess of the voice). Other PQs, especially breathiness with a correlation of 0.87, are easier for non-experts to hear and rate. In terms of agreement with experts, non-experts achieve a surprising level of performance. 

\begin{figure}[t]
     \centering
         \centering
         \includegraphics[width=0.45\textwidth]{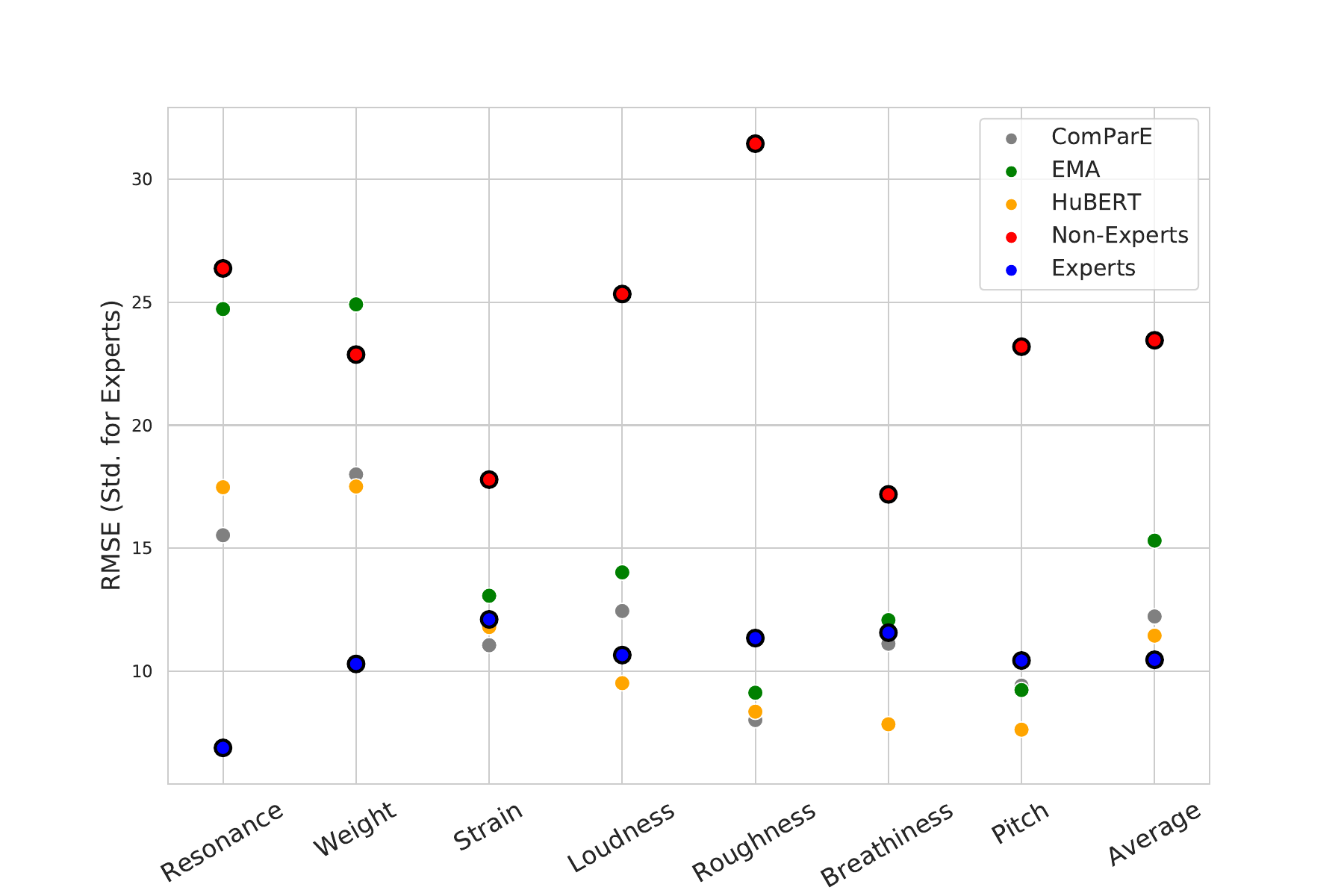}

        \caption{Test RMSE of various rating methods when compared with the average expert rating for each perceptual quality. Standard Deviation reported for expert ratings.}
        \label{fig:stds}
\end{figure}

While the correlation is remarkably high between non-experts and experts, RMSE, as reported in Figure \ref{fig:stds}, suggests a high-level of deviance between non-experts and experts. While experts have an average standard deviation amongst themselves of 10.47, average non-expert RMSE with expert ratings is 23.45, slightly over twice that of experts. 

The RMSE between non-expert and expert ratings suggests a potential flaw in rating ability of non-experts, but visualizing the average non-expert ratings vs. average expert ratings, reported in Figure \ref{fig:ne_vs_e} reveals non-expert biases across PQs. For CAPE-V PQs, which are all measures of deviance, non-experts consistently overrate speech clips as having higher levels of deviance. But, for those clips that experts label as having high levels of deviance, non-experts will always label as having high-levels of deviance as well. Regarding the gendered PQs of resonance and weight, we see higher levels of deviation between non-expert and expert ratings, but similar trends hold. Non-experts appropriately label voices with high-values of a perceptual quality with a high-value, and vice-versa. 

Further training or better explanation is most likely required to improve non-expert and expert agreement on the gendered PQs. These results, however, have shown that the ability of non-experts to hear perceptual qualities with minimal prompting is remarkably high and the promise of collecting mass perceptual quality data is well within reach.

% Given the trends seen across non-expert ratings of all perceptual qualities, we hypothesize that bias correction for non-expert ratings could greatly improve the agreement between non-expert and expert ratings. We leave such exploration for future work. 

% Describe the AMT experiments. Lowest correlation and highest correlation. Show graphs of expert avg. vs. non-expert avg. Conclude that non-experts can hear the perceptual qualities.

\section{Can Objective Features Predict Subjective Qualities?}

While prior work has demonstrated the capability of predicting perceptual qualities directly from waveform and mel-spectrograms \cite{hidaka2022deepgrbas, fujimura2022classification}, in this section, we explore the universality of perceptual qualities across all various representations of speech: acoustic, physical, and self-supervised.

% \begin{table*}[t]
%     \centering
%     % \begin{tabular}{*{8}{c}}
%     \begin{tabular}{|c|c|c|c|c|c|c|c||c|}
%         \hline
%         \textbf{Rater} & \textbf{Resonance} & \textbf{Weight} & \textbf{Strain} & \textbf{Loudness} & \textbf{Roughness} & \textbf{Breathiness} & \textbf{Pitch} & \textbf{Average}  \\
%         \hline
%          ComParE & 15.53 & 18.00 & 11.06 & 12.45 & 8.02 & 11.12 & 9.42 & 12.23 \\
%         \hline
%          EMA & 24.72 & 24.91 & 13.07 & 14.02 & 9.13 & 12.08 & 9.24 & 15.31 \\
%         \hline
%          HuBERT & 17.48 & 17.51 & 11.80 & 9.52 & 8.36 & 7.85 & 7.63 & 11.45 \\
%         \hline
%          Non-Experts & 26.37 & 22.87 & 17.79 & 25.33 & 31.44 & 17.19 & 23.19 & 23.45 \\
%          \hline
%          Experts & 6.89 & 10.30 & 12.11 & 10.66 & 11.35 & 11.57 & 10.44 & 10.47 \\

%          \hline
%     \end{tabular}
%     \caption{RMSE of various rating methods when compared with the average expert rating for each perceptual quality. For resonance and weight, results are reported against only one rater. RMSE for models is reported on a held-out test set. Standard Deviation reported for expert ratings.}
%     \label{tab:stds}
% \end{table*}

\subsection{Random Forest Regression}

Given the highly non-linear relationship and high-dimensionality of various representations, we trained random forest regressors on a 60-20-20 train-validation-test split on PVQD+, finding that linear models such as Lasso resulted in lower performance. Hyperparameters for random forests were fine-tuned on the 20\% validaiton set.

\subsection{Feature Sets}

 Three feature sets are used in the random forest regression models. For the acoustic features, we use ComParE 2016 feature set, which consists of a combination of functionals computed over prosodic, spectral, and sound quality-based features \cite{weninger2013acoustics}. For the physical features, we use the Electromagnetic articulography (EMA) representation of speech, which tracks movement of articulators with midsagittal x, y coordinates of jaw, lips, and tongue positions \cite{wu2023speakerindependent}. For the self-supervised features, we use the 7th layer of a pre-trained HuBERT model, which distills speech into a 1024-dimension learned representation via training on masked audio samples \cite{hsu2021hubert}. 

% \noindent\textbf{Acoustic Features}\label{rf_model}: ComParE 2016 feature set, which consists of a combination of functionals computed over prosodic, spectral, and sound quality-based features \cite{weninger2013acoustics}.

% \noindent\textbf{Physical Features}: Electromagnetic articulography (EMA) representation of speech, which tracks movement of articulators with midsagittal x, y coordinates of jaw, lips, and tongue positions \cite{wu2023speakerindependent}.

% \noindent\textbf{Self-Supervised Features}: 7th layer of a pre-trained HuBERT model, which distills speech into a 1024-dimension learned representation via training on masked audio samples \cite{hsu2021hubert}. 

\subsection{Results}

The test RMSE for the above representations are reported in Figure \ref{fig:stds}. Due to the lack of data, RMSE for resonance and weight is reported against only one voice teacher's ratings. 

Across all three representations, several trends become clear. On average, all random forests across feature sets predict expert ratings with lower error than non-expert humans. Additionally, for CAPE-V PQs, ComParE and HuBERT features both achieve RMSE lower than inter-expert standard deviation, with the exception of Loudness for ComParE. EMA sees lower RMSE for two of the five CAPE-V PQs,  and similar performance for both breathiness and strain. 

While the models performed well for most of the CAPE-V PQs, the models failed to achieve similar performance on the gendered PQs. Considering the reliance of resonance and weight on laryngeal and source information (Section \ref{sec:pvqd_plus}), EMA's lack of such information justifies its poor performance. Given the excellent performance of ComParE and HuBERT features on the CAPE-V perceptual qualities, however, the performance drop for resonance and weight is surprising. While the performance is still better than that of non-experts (save for EMA and weight), we expect that this performance drop is in part due to the lack of labels for the entirety of the PVQD dataset. 

% Investigating the correlations between ComParE features reveals perhaps one reason why the 

% \robbie{Discuss correlations or just discuss performance? Running out of space} 

\section{Discussion and Future Work}\label{sec:disc}

Perceptual qualities are a perceivable and predictable representation of speaker identity. With minimal examples, non-experts can label perceptual qualities with remarkable correlation to expert labels. Across multiple speech representations, perceptual qualities are predictable, and, for CAPE-V PQs, at an error that is lower than inter-expert variation. 

While the current work points at the ability of perceptual qualities to capture information about speaker identity, future work is needed to explore the limits of perceptual qualities. To what extent can PQ-based representations uniquely identify a voice? Can perceptual qualities be used to guide speech synthesis or voice conversion? Many questions on perceptual quality's applications remain, but the perceivability and ubiquity of perceptual qualities promises to bring unseen interpretability and flexibility to speech processing systems.

% \section{Conclusion}\label{sec:conc}

\bibliographystyle{IEEEbib}
\bibliography{main}

\end{document}